\documentclass[english]{article}
\usepackage[T1]{fontenc}
\usepackage[latin9]{inputenc}
\usepackage{amsmath}
\usepackage{amssymb}
\usepackage{algorithm}
\usepackage{algpseudocode}
\usepackage{color}
\usepackage{graphicx}

\newcommand\blfootnote[1]{%
  \begingroup
  \renewcommand\thefootnote{}\footnote{#1}%
  \addtocounter{footnote}{-1}%
  \endgroup
}

\begin{document}
\title{Real-Space Renormalization group for spin glasses}\label{sec:RS_RG}
\author{Maria Chiara Angelini$^1$}
\date{\footnotesize $^1$Dipartimento di Fisica, Sapienza Universit\`a di Roma, P.le Aldo Moro 5, 00185 Rome, Italy and 
Istituto Nazionale di Fisica Nucleare, Sezione di Roma I, P.le A. Moro 5, 00185 Rome, Italy}

\maketitle

\begin{abstract}
While in the fully-connected limit the solution of the spin-glass model is known, 
with the existence of a complex transition on a critical line in the temperature-external field phase diagram, 
in finite dimensions we don't know if a transition is present or not and, if present, 
if its nature is the same as the one in infinite dimensions.     
This work contains a review of the real-space Renormalization Group methods that have been applied to Spin Glasses, 
highlighting both their point of strength and weakness in characterizing the finite dimensional behaviour of the model.
\end{abstract}

\blfootnote{To appear as a contribution to the edited volume "Spin Glass Theory and Far Beyond - Replica Symmetry Breaking after 40 Years", P. Charbonneau, E. Marinari, M. Mezard, G. Parisi, F. Ricci-Tersenghi, G. Sicuro and F. Zamponi (World Scientific)}

Spin glasses (SGs) are the prototype of \textit{strongly disordered systems}.
While in infinite $d$ the solution of the model is known \cite{parisi1980sequence, parisi1980order}, with the existence of a complex SG transition on a critical line in the temperature-external field phase diagram, in finite $d$ we don't know if a transition is present or not and, if present, if its nature is the same as the one in infinite $d$. 

A standard perturbative renormalization Group (RG) computation at one loop for the spin-glass in a field finds no suitable fixed point (FP) below the upper critical dimension $d_u=6$ describing the low-temperature phase \cite{bray1980renormalisation,pimentel2002spin}. On the other hand, the perturbative expansion up to the second order finds a strong-coupling FP \cite{charbonneau2017nontrivial}. However, this new FP is in a way "nonperturbative" as it cannot be reached continuously from the mean-field (MF) one just by lowering the dimension. Being the perturbative analysis in the strong-coupling regime uncontrolled, the existence and relevance of this new FP cannot be stated just with the methods of Ref. \cite{charbonneau2017nontrivial}.

The use of real-space RG methods could then seem a natural choice if we are looking for nonperturbative FPs in finite dimensions: 
Real-space RG methods are non-perturbative by construction. In this section we will try to review the real-space RG methods that have been applied to SGs, highlighting both their point of strength and weakness. 

\section{The Migdal-Kadanoff Renormalization Group method}
\begin{figure}
  \centering
  \includegraphics[width=.5\linewidth]{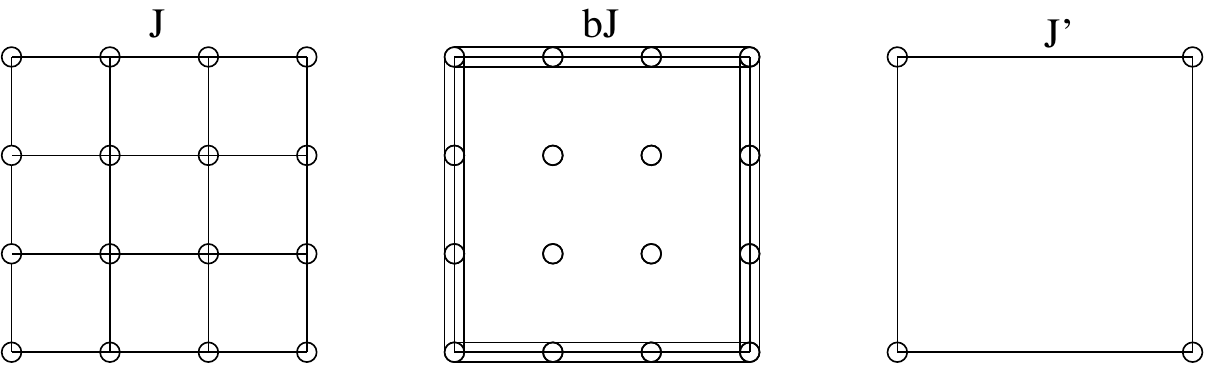}  
  \caption{Basic step of the Migdal-Kadanoff bond moving procedure to renormalize an hypercubic lattice in $d=2$ dimensions.}
  \label{Fig:MK}
\end{figure}
\begin{figure}
  \centering
  \includegraphics[width=.5\linewidth]{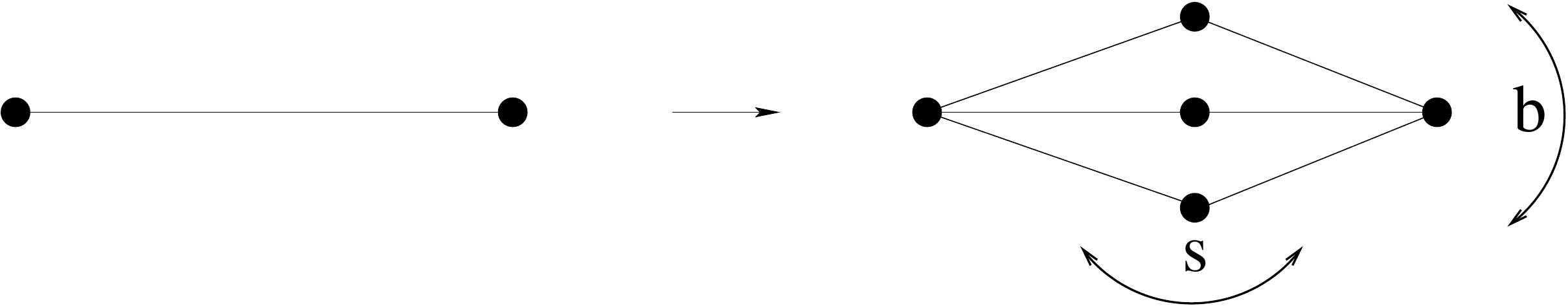}  
  \caption{Basic step of the iterative procedure to generate a Hierarchical diamond lattice. MK RG is exact on lattices of this type.}
  \label{Fig:HL}
\end{figure}

The real-space RG can be viewed as a decimation procedure that starts with a larger system and reduces it to a smaller system, in such a way to preserve, or to scale appropriately, some important physical observables. Such a decimation induces an RG transformation on the system couplings, and the study of such a transformation allows one to identify critical points and critical exponents. the partition function is thus evaluated by steps. In each step a block of spins $\{\sigma\}$, described by the Hamiltonian $H(\{\sigma\})$ with couplings $\{J\}$,
is replaced by an equivalent system with fewer spins $\{\sigma'\}$ and Hamiltonian $H'(\{\sigma'\})$, with
new, renormalized couplings $\{J'\}$ in such a way that the partition function of the original and the renormalized systems are the same.
While this procedure can be carried out explicitly in dimension $d=1$ because the Hamiltonian remains of the same form after the reduction of the degrees of freedom, in dimension $d>1$ new couplings terms arise between distant spins, and the block-spin renormalization cannot be carried out exactly. The Migdal-Kadanoff (MK) approximation is a way to overcome the problem of proliferation of the couplings \cite{kadanoff1975variational,migdal1976phase}. Once the spins in the lattice are divided into blocks, all the
couplings internal to the blocks are moved to the spins at the edges of the blocks. At this point, an exact decimation of the spins at the edges, except those on the corners, is performed. This procedure is illustrated in Fig. \ref{Fig:MK}.
One can demonstrate that the free energy of the system after the bond-moving procedure, is a lower bound to the free energy of the original system.
The MK procedure applied to a hypercubic lattice in $d$ dimensions consists in replacing it with a hierarchical diamond one, for which the MK RG is exact \cite{berker1979renormalisation}. Hierarchical diamond lattices (HL) are generated iteratively. The procedure starts at step $G=0$ with two spins connected by a single link.
At each step $G$, for each link of step $G-1$, $b$ parallel
branches, made of $s$ bonds in series each, are added,
creating $b\cdot (s-1)$ new spins. The first step of this iterative procedure is shown in Fig. \ref{Fig:HL}. The relationship between the dimension of the hypercubic lattice and the number of branches and bonds in the associated hierarchical lattice is
$d=1+\ln(b)/\ln(s)$. 
The RG procedure is exactly the opposite of
the iterative procedure to construct the HL. For instance, in
step 1, the $b\cdot (s-1)$ spins generated at the last level are integrated
out, generating new effective couplings and fields between the
remaining spins \footnote{In the following we will assume $s=2$ because this is the chosen value in all the considered works}. 
Particular care should be taken when fields are involved \cite{drossel2000spin}.
Despite its simplicity, the MK renormalization can capture highly nontrivial features of the analyzed models, for example,
it accurately describes the zero-temperature FP of the random
field Ising model \cite{cao1993migdal}. On the
other hand, it becomes less quantitatively accurate in high
dimensions and sometimes even fails qualitatively \cite{antenucci2014critical}.

\begin{figure}
\begin{center}
  \includegraphics[width=2in]{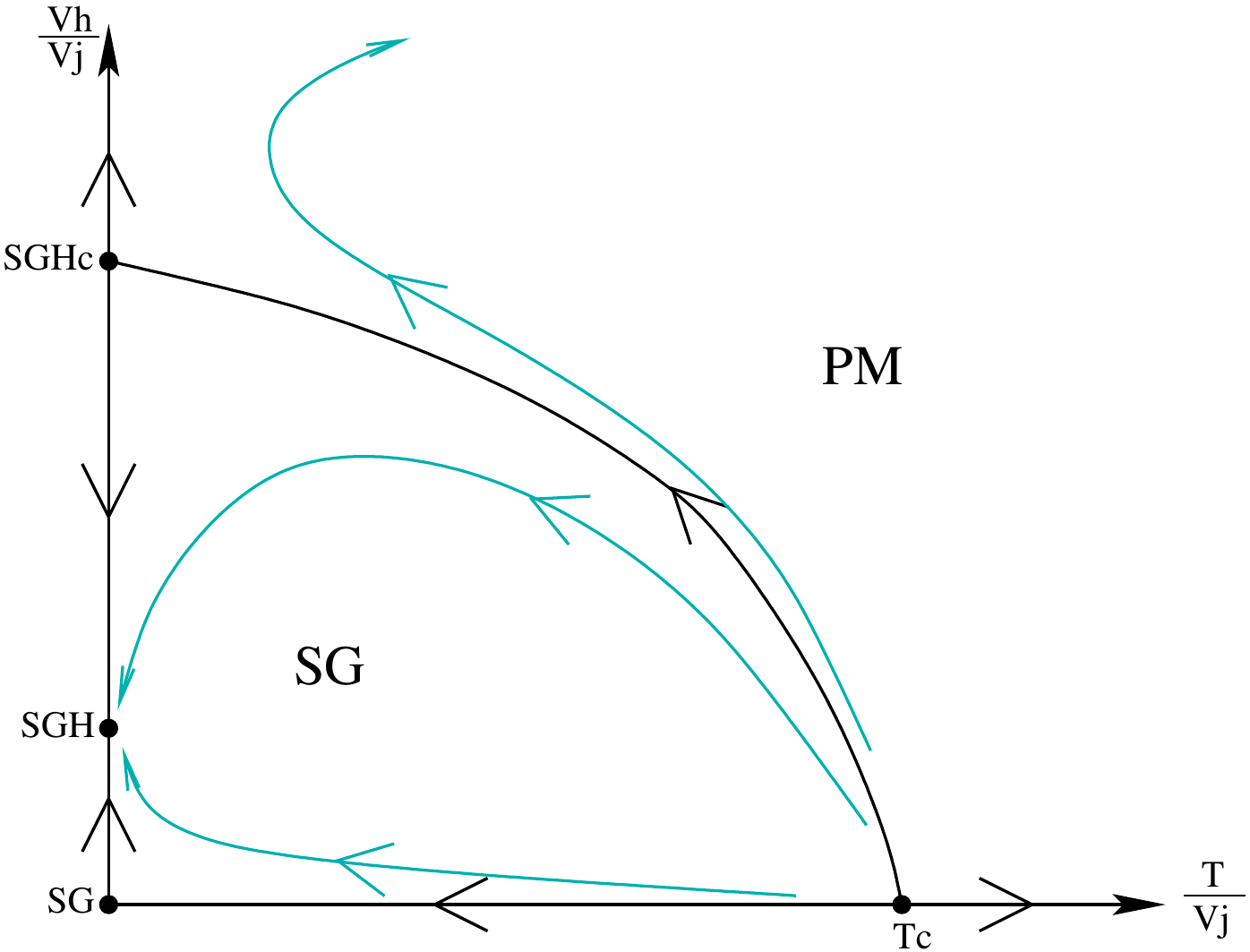}
  \caption{Renormalization flow in the plane $(\frac{T}{v_j}-\frac{v_h}{v_j})$ obtained from MK RG for $d>8$. 
Reprinted figure with permission from [Maria Chiara Angelini and Giulio Biroli
Phys. Rev. Lett. 114, 095701, 2015] Copyright (2015) by the American Physical Society.}
  \label{Fig:MKdiagram}
  \end{center}
\end{figure}

The phase diagram for SG without fields obtained through MK RG is depicted in ref. \cite{southern1977real}.
In this case, the model displays a phase transition from a paramagnetic (PM) to an SG
phase at $T_c(b)$. Starting at $T>T_c$, the renormalized variance of the coupling distribution $v_J$ decreases, flowing towards the paramagnetic fixed point $\frac{T}{v_j}=\infty$. 
On the other hand, starting at $T<T_c$, the renormalized variance of the coupling distribution increases towards a zero temperature FP associated with the SG phase $\frac{T}{v_j}=0$. At the SG FP, the renormalized variance of
the couplings after $n$ iterations grows as $v_J^{(n)}\propto (2^n)^{\theta}$. The dependence of the exponent $\theta$ on the effective dimension is well described by $\theta(d)= (d-2.5)/2$, which is consistent with the lower critical dimension $d_L=2.5$ obtained by numerical and theoretical arguments \cite{Franz1994DL,Boettcher2005DL,maiorano2018support}.

In the spin-glass phase, single renormalization-group trajectories are chaotic \cite{mckay1982spin}, and renormalized couplings display a chaotic temperature dependence \cite{nifle1992new}.

One can then look at how this picture is modified by the introduction of random fields. Let us suppose that the original fields are extracted from a Gaussian distribution of zero mean and variance $v_{h}$.
One can show that, at any dimension (any $b$), the zero temperature SG FP $\frac{T}{v_j}=0$ becomes unstable, the external field corresponding to a relevant
perturbation. For $d$ small enough there is no other stable FP associated to the SG phase with field \cite{drossel2000spin}: the transition seems to be destroyed by the field. However things change in higher dimension: The renormalization flow projected in the plane $(\frac{T}{v_j},\frac{v_h}{v_j})$ for $d>8$ is qualitatively shown in Fig. \ref{Fig:MKdiagram} \cite{angelini2015spin}:
 Even though the SG-FP is again unstable in presence of an external field,  the system flows toward a new zero-temperature stable fixed point, SGH,  which rules the behavior of the SG phase in a field.
At high temperature and/or for strong fields the system flows to the PM FP $(\frac{T}{v_j},\frac{v_h}{v_j})=(\infty,\infty)$; thus, there is necessarily an unstable FP, SGH$_c$, separating the PM and the SGH ones. This is also at zero temperature and governs the transition of SGs in a field. The fact that the critical FP point is a zero temperature one, implies that there is a third independent critical exponent, in addition to the usual two exponents associated with finite T FPs. Again, one can compute the new exponent $\theta_{c}$, looking at how the variance of the couplings increases at the SGH$_c$ FP. The other consequences of a $T=0$ FP is that, while correlation functions associated to thermal fluctuations decay as $G_{thermal}(r)\propto \frac{1}{r^{d-2+\eta}}$, correlation functions associated to disordered fluctuations decay as $G_{disorder}(r)\propto \frac{1}{r^{d-2+\eta-\theta_c}}$ \cite{bray1985scaling}. The RG picture obtained from MK is thus deeply different from the standard MF theory, which predicts a Gaussian FP in high dimensions that is not a zero-temperature one (and thus thermal and disordered correlation functions that decay with the same exponent). Indeed the Gaussian FP exactly describes the Fully-connected SK model that does not have a transition when $T=0$, being always in the RSB phase. 
For $d\to\infty$, the transition found through MK RG loses its zero-temperature character since $\theta_c\to 0$.
The MK RG predicts a lower critical dimension $d_L=8$ below which it does not find a stable fixed point when the field is present. The same MK RG method has also been applied to models of glasses for which the microscopic degrees of freedom can take $q$ values. While the RG flow is similar to Fig. \ref{Fig:MKdiagram}, with the presence of a critical line ending on a $T=0$ critical FP, and the low-temperature phase governed by another $T=0$ fixed point, the lower critical dimension decreases when increasing $q$. While $d_L(q=2)=8$, $d_L(q=\infty)\simeq 4$ \cite{angelini2017PNAS,angelini2017real}.

The curse of the Migdal-Kadanoff RG method is that basically, it assumes from the outset that the system is replica symmetric: as shown analytically by Gardner \cite{Gardner1984}, it cannot include RSB, reducing the operative space to a finite, discrete space: if one wants to understand if the finite-dimensional world has RSB, then one needs other RG methods. 

\section{The ensemble RG method}

The MK approach is a transformation that at each step maps a single sample of size $N$ to a smaller system. Given an ensemble of systems of size $N$, a transformation is applied to each of them to obtain an ensemble of smaller systems. However, a different approach is possible, establishing a direct mapping between the entire probability distributions of couplings in larger and smaller systems, such that the average over such distributions of important observables remains the same. Obviously, for models for which the RG transformation is exact, the two approaches should provide the same answer, but when approximations are made, the latter could lead to better results: in models with strong disorder (like SGs), sample-to-sample fluctuations may dominate thermal ones.
Following the latter approach, in Ref. \cite{angelini2013ensemble} the Ensemble RG (ERG), is proposed.
In principle, the ERG can be applied to any disordered system. However, it has been applied so far only to the hierarchical model (HM), which is a particular one-dimensional long-range model, whose Hamiltonian for $N=2^n$ spins can be constructed iteratively in the following way:

\footnotesize

\begin{align*}
H_{n+1}(s_1,...,s_{2^{n+1}})=H_{n}(s_1,...,s_{2^{n}})+H_{n}(s_{2^n+1},...,s_{2^{n+1}})+c^{n+1}\sum_{i<j=1}^{2^{n+1}}J_{ij}s_i s_j+cost 
\end{align*}

\normalsize
In practice, $H_n$ is the sum of interactions at $n$ different levels.
The hierarchical model was introduced by Dyson in its ferromagnetic version \cite{dyson1969existence} (for a review see \cite{meurice2007nonlinear}), and reproposed in its SG version in Ref. \cite{franz2009overlap}.
By properly tuning the factor
$c$ controlling how fast the couplings' intensity decays with
distance, the HM can emulate a $d$-dimensional short-range
(SR) model: $c\simeq2^{-1-2/d}$ for the ferromagnetic model, $c\simeq2^{(-1-2/d)/2}$ for the SG version (more refined relations can be written  \cite{angelini2014relations, banos2012correspondence}).

If an HM is decimated by a standard block-spin transformation, the
new Hamiltonian does not contain any multispin terms (at
variance to what happens on finite-dimensional lattices).
So, considering only pairwise interactions in the RG is not
an approximation for the HM. 

In the ERG, couplings are assumed to remain independent, but they can have a different probability
distribution $P_k(J)$ at each level $k\in\{1,2,...,n\}$. Each coupling distribution is parametrized by $K$ numbers otherwise the RG for the entire coupling distributions would become untractable: for SG with a field, one can assume the distribution of couplings and fields to be two independent Gaussians, and one has just $K=2$ parameters that are the variance of couplings and fields. The ERG for an ensemble of systems
with $n$ levels works as follows:
\begin{enumerate}
    \item Compute $(n - 1)K$ observables $<O_j>$, $j \in\{K + 1, ... ,Kn\}$ in the larger systems extracted from the original coupling distribution.
\item Determine the new $(n -1)K$ parameters of the
$P'$ distributions by requiring that $<O'_i>_{P'}=<O_{i+K}>_P$ for any
$i \in {1,2,...,(n -1)K}$.

\item Build a new ensemble of systems of the
original size. They are constructed joining with random
couplings extracted from the original distribution $P_n(J)$ two
smaller systems with couplings extracted from $P'_k(J')$, $k\in\{1,2,...,(n -1)\}$ found at step (2).
\end{enumerate}
Primed quantities refer to the smaller systems. The first two steps are the true renormalization steps, while the last step is required to obtain a final system size that allows for iterating the method
until convergence. 
The observables used to fix the variances in the SG ERG are normalized SG correlations at different levels.
In Ref. \cite{angelini2013ensemble} the case of SG without field is analyzed, finding an SG transition below a critical temperature also for effective dimension $d\simeq 3$.
The reliability of the method
has been tested by comparing the values of critical temperatures
and critical exponents with those obtained in MC simulations.
The ERG method can reproduce the correct behavior
of the $\nu$ exponent, which, for long-range systems, shows a minimum at the upper critical value of $c$: in this way the ERG method correctly identifies the upper critical dimension of SGs in zero field.
The ERG was applied to SG with a field in ref. \cite{angelini2015spin}. In this case, the results are in perfect agreement with what was found by MK RG: the qualitative phase diagram is again the one in Fig. \ref{Fig:MKdiagram}, and below $d_L\simeq 8$ the ERG method is not able to identify an SG phase when the field is present.

\section{The Strong disorder RG}

In this section, we will review the Strong Disorder RG (SDRG) as introduced in ref. \cite{monthus2015fractal}: an RG at zero temperature that permits the construction of an approximated ground state of an SG system.

For each spin $S_i$, its local field
$h^{loc}_i = \sum_{j} J_{ij}  S_j$
is considered. Once its largest coupling in absolute value is computed, corresponding to some index $j_{max}(i)$, $\max_{j} ( \vert J_{ij}  \vert  ) \equiv \vert J_{i,j_{max}(i)}  \vert$, one would like to identify the spins for which the local field
\begin{eqnarray*}
h^{loc}_i && =  J_{i,j_{max}(i)}   S_{j_{max}(i)}+ \sum_{j \ne j_{max}(i)}  J_{ij} S_j
\label{hloci2}
\end{eqnarray*}
 is dominated by the first term. The second term could be approximated by a sum of random 
terms of absolute values $J_{ij}$ and of random signs, i.e. it is reasonable to look at
\begin{eqnarray*}
\Omega_i && \equiv \vert J_{i,j_{max}(i)}  \vert - \sqrt{ \sum_{j \ne j_{max}(i)}  \vert J_{ij}  \vert^2  }
\label{omegaii}
\end{eqnarray*} 
as an indicator of the relative dominance of the maximal coupling in the local field.
The Strong Disorder RG procedure based on the variable $\Omega_i$, introduced in ref.\cite{monthus2015fractal}, is defined by the following elementary decimation steps:
\begin{enumerate}
\item For each spin $i$, compute the associated variable $\Omega_i$
\item Find the spin $i_0$ with the maximal $\Omega_i$
\item Eliminate the spin $S_{i_0}$, fixing it to the value: 
\begin{eqnarray*}
S_{{i_0}} =  S_{j_{max}({i_0})} {\rm sgn} (J_{{i_0} j_{max}({i_0})})
\label{elimsi0}
\end{eqnarray*}
\item Transfer all its couplings $J_{i_0,j} $ with $j \ne j_{max}(i_0)$ 
to the spin $S_{j_{max}(i_0)}$ via the renormalization rule
\begin{eqnarray}
J^R_{j_{max}(i_0),j} = J_{j_{max}(i_0),j}+ J_{i_0,j}  {\rm sgn} (J_{i_0 j_{max}(i_0)}).
\label{rulejr}
\end{eqnarray}

\end{enumerate}

The procedure is repeated $N-1$ times; at the end, only a single spin $S_{last}$ is left:
the two values $S_{last}=\pm 1$ label the two ground states related by a global flip of all the spins.
From the choice $S_{last}=+1$, one may reconstruct all the values of the
decimated spins via the rule of Eq. (\ref{elimsi0}), and can thus approximate the energy of the ground-state.

In ref. \cite{wang2018fractal}, the strong disorder RG has been used to make predictions on the validity of Droplet Picture (DP) or Replica Symmetry Breaking (RSB) theory depending on the dimension $d$. In the DP the low-temperature phase is replica symmetric and its properties are determined by the excitation of droplets whose free-energy cost on a length scale $L$ goes as $L^\theta$ and which have fractal dimension $d_s < d$. 
In the RSB picture there exist system-size excitations which have a free-energy cost of O(1) and which
are space-filling, i.e., have $d_s = d$. Thus, by investigating the
value of $d_s$ of interfaces in the low-temperature phase, it is
possible to determine whether the low-temperature state is best
described by RSB or DP. 

$\theta$ and $d_s$ can be found by considering, in each disordered sample, the two ground states associated with two different boundary conditions, for instance, Periodic (P) and
Anti-Periodic (AP) (One can change from periodic to antiperiodic boundary conditions by flipping the sign of the bonds crossing a hyperplane of
the lattice). The difference between the two ground states defines a system-size Domain-Wall. From the scaling of its energy one extracts $\theta$, and from the scaling of its surface one extracts $d_s$.

In ref. \cite{monthus2015fractal}, $\theta$ and $d_s$ are computed by SDRG for $d=2,3$. While the values of $d_s$ by SDRG are in good agreement with the ones obtained by numerical methods both in $d=2$ \cite{khoshbakht2018domain}, and in $d=3$ \cite{wang2017number}, $\theta$ is not well captured by SDRG, giving $\theta(d=2)\simeq 0$. SDRG thus seems to behave in the opposite way  w.r.t. the MKRG, that correctly predicts the value of $\theta$, and misses the value of $d_s$ which is fixed to the
trivial value $d_s^{MK}=d-1$.

In ref. \cite{wang2018fractal}, the value for $d_s$ is computed by SDRG, and complemented by using a greedy algorithm, for dimensions up to $d=6$: the two estimates appear to merge and give $d_s = d$ in $d = 6$, thus suggesting that RSB could be valid above $d=6$, while DP could describe the model for $d<6$.

The main problem of SDRG is that it seems to be accurate in the early stages of the RG process where
there exist spins with positive (and large) $\Omega_i$. All $\Omega_i$ turn negative for
the later stages of the iteration procedure, indicating that the
SDRG is failing. As suggested by Monthus \cite{monthus2015fractal}, it could be that the fractal dimension $d_s$ is dominated by the early stages of the iteration, which
correspond to long length scales, and for this reason, the SDGR correctly captures its value.

\section{The M-Layer expansion around the Bethe lattice solution}
 Finally, we mention a quite recent expansion around a
different soluble MF model: the Bethe lattice (BL) (with BL we will call
a random regular graph of finite connectivity $z$). A model on a BL
is essentially MF because of the local tree structure of the lattice: The probability of loops of a finite
length goes to zero in the thermodynamic limit, and for this reason, the probability distribution of a
spin is independent of the probability of a nearest neighboring spin if the direct edge between them
is cut. 
The idea of an expansion around the BL was originally introduced by Efetov \cite{EFETOV1990}
and successively revived by different authors \cite{Slanina2006,Sacksteder2007}.
In ref. \cite{altieri2017loop}, it was formalized through the $M$-layer construction: one introduces $M$ copies of the original finite-dimensional lattice and generates a new lattice through a local random rewiring of the links. In the $M\to\infty$ limit, the resulting $M$-layer lattice looks locally like a BL, with a tree-like local structure without loops of finite length.
On the opposite side, for $M=1$ one recovers the original lattice.
Introducing the small parameter $1/M$, one can perform an expansion for a generic multi-point observable: The critical series is expressed as a sum of topological Feynman diagrams with the same numerical pre-factors they have in field theories. The only difference is that the contribution of a given diagram must not be evaluated as usual associating bare propagators to its lines; instead, one needs to compute the observable on the corresponding topological loop diagram, thought as manually inserted in a BL. At leading order, no spatial loops are considered and one recovers the BL solution, while spatial loops become more and more important in finite-dimensional systems lowering the dimension: for this reason they are present at higher orders in the BL expansion.
The BL expansion has a perturbative nature, a very useful property to keep computations under control. Moreover, it permits following the well-known path traced by standard perturbative RG. However, the BL-RG includes also non-perturbative features w.r.t. the standard expansion: The BL solution is exact in one dimension, thus including the resummation of all the non-perturbative effects. Finite connectivity is already encoded in the 0th order of the expansion, and, as a direct consequence,
also important properties such as local fluctuation of observables and heterogeneity, at variance with
the expansion around the fully-connected (FC) MF solution where they could only be seen as non-perturbative effects. In Ref. \cite{angelini2018one} the $M$-layer expansion around BL is performed for the SG with a field in the limit of large connectivity $z\to\infty$ at positive temperature, recovering the same results of the standard expansion \cite{bray1980renormalisation,pimentel2002spin}.

In precedent sections, we have seen that non-perturbative RG schemes such as MK or ERG, find in high enough dimensions a critical zero-temperature fixed point for the SG with field. In the FC model, the transition line in the $(T-h)$ temperature-field plane diverges at $T = 0$: there are no zero-temperature fixed points around which one could expand. On the contrary, on the BL at
$T = 0$ there is a transition at a finite field $h_c$\cite{parisi2014diluted}, around which one can perform an expansion using the $M$-layer formalism. While setting the temperature straight to 0 is impossible in the Lagrangian approach of the FC expansion, $T = 0$ computations can be easily performed in the context of the BL expansion \cite{angelini2020loop}.
In Ref. \cite{angelini2022unexpected} the 0-loop two-point correlation functions and the first 1-loop corrections are computed exactly at $T=0$ and finite connectivity: loop corrections are not
negligible for $d<d_u^{BL}=8$. The upper critical dimension predicted by the BL expansion is different from the one predicted by standard field theory $d_u=6$. Moreover, if one
takes the limit $T\to 0$ of the BL expansion performed at large connectivity and $T\neq 0$ in ref. \cite{angelini2018one}, the 1-loop correction results to be of the standard form. Finite connectivity is thus a crucial ingredient in the computation and the limits $z\to \infty$
and $T\to0$ do not commute. 

At this point, one should compute three-point correlation functions at zero and one loop, associated with the cubic vertex, inside the BL expansion, and see if, by standard RG field theoretical methods, one can find a non-trivial FP of the RG equations, below the upper critical dimension $d_u^{BL}$.
This program is currently underway.

\bibliography{bibfile}

\begin{thebibliography}{10}

\bibitem{parisi1980sequence}
Giorgio Parisi.
\newblock A sequence of approximated solutions to the sk model for spin
  glasses.
\newblock {\em J.\ Phys.\ A}, 13(4):L115, 1980.

\bibitem{parisi1980order}
Giorgio Parisi.
\newblock The order parameter for spin glasses: a function on the interval 0-1.
\newblock {\em J.\ Phys.\ A}, 13(3):1101, 1980.

\bibitem{bray1980renormalisation}
AJ~Bray and SA~Roberts.
\newblock Renormalisation-group approach to the spin glass transition in finite
  magnetic fields.
\newblock {\em Journal of Physics C: Solid State Physics}, 13(29):5405, 1980.

\bibitem{pimentel2002spin}
IR~Pimentel, T~Temesv{\'a}ri, and C~De~Dominicis.
\newblock Spin-glass transition in a magnetic field: A renormalization group
  study.
\newblock {\em Physical Review B}, 65(22):224420, 2002.

\bibitem{charbonneau2017nontrivial}
Patrick Charbonneau and Sho Yaida.
\newblock Nontrivial critical fixed point for replica-symmetry-breaking
  transitions.
\newblock {\em Physical review letters}, 118(21):215701, 2017.

\bibitem{kadanoff1975variational}
Leo~P Kadanoff.
\newblock Variational principles and approximate renormalization group
  calculations.
\newblock {\em Physical Review Letters}, 34(16):1005, 1975.

\bibitem{migdal1976phase}
Alexander~A Migdal.
\newblock Phase transitions in gauge and spin-lattice systems.
\newblock {\em Soviet Journal of Experimental and Theoretical Physics}, 42:743,
  1976.

\bibitem{berker1979renormalisation}
As~N Berker and S~Ostlund.
\newblock Renormalisation-group calculations of finite systems: order parameter
  and specific heat for epitaxial ordering.
\newblock {\em Journal of Physics C: Solid State Physics}, 12(22):4961, 1979.

\bibitem{drossel2000spin}
Barbara Drossel, Hemant Bokil, and MA~Moore.
\newblock Spin glasses without time-reversal symmetry and the absence of a
  genuine structural glass transition.
\newblock {\em Physical Review E}, 62(6):7690, 2000.

\bibitem{cao1993migdal}
MS~Cao and J~Machta.
\newblock Migdal-kadanoff study of the random-field ising model.
\newblock {\em Physical Review B}, 48(5):3177, 1993.

\bibitem{antenucci2014critical}
F~Antenucci, A~Crisanti, and L~Leuzzi.
\newblock Critical study of hierarchical lattice renormalization group in
  magnetic ordered and quenched disordered systems: Ising and
  blume--emery--griffiths models.
\newblock {\em Journal of Statistical Physics}, 155(5):909--931, 2014.

\bibitem{southern1977real}
BW~Southern and AP~Young.
\newblock Real space rescaling study of spin glass behaviour in three
  dimensions.
\newblock {\em Journal of Physics C: Solid State Physics}, 10(12):2179, 1977.

\bibitem{Franz1994DL}
{S. Franz}, {G. Parisi}, and {M.A. Virasoro}.
\newblock Interfaces and louver critical dimension in a spin glass model.
\newblock {\em J. Phys. I France}, 4(11):1657--1667, 1994.

\bibitem{Boettcher2005DL}
Stefan Boettcher.
\newblock Stiffness of the edwards-anderson model in all dimensions.
\newblock {\em Phys. Rev. Lett.}, 95:197205, Nov 2005.

\bibitem{maiorano2018support}
Andrea Maiorano and Giorgio Parisi.
\newblock Support for the value 5/2 for the spin glass lower critical dimension
  at zero magnetic field.
\newblock {\em Proceedings of the National Academy of Sciences},
  115(20):5129--5134, 2018.

\bibitem{mckay1982spin}
Susan~R McKay, A~Nihat Berker, and Scott Kirkpatrick.
\newblock Spin-glass behavior in frustrated ising models with chaotic
  renormalization-group trajectories.
\newblock {\em Physical Review Letters}, 48(11):767, 1982.

\bibitem{nifle1992new}
M~Nifle and HJ~Hilhorst.
\newblock New critical-point exponent and new scaling laws for short-range
  ising spin glasses.
\newblock {\em Physical review letters}, 68(20):2992, 1992.

\bibitem{angelini2015spin}
Maria~Chiara Angelini and Giulio Biroli.
\newblock Spin glass in a field: A new zero-temperature fixed point in finite
  dimensions.
\newblock {\em Phys.\ Rev.\ Lett.}, 114(9):095701, 2015.

\bibitem{bray1985scaling}
AJ~Bray and MA~Moore.
\newblock Scaling theory of the random-field ising model.
\newblock {\em Journal of Physics C: Solid State Physics}, 18(28):L927, 1985.

\bibitem{angelini2017PNAS}
Maria~Chiara Angelini and Giulio Biroli.
\newblock Real space renormalization group theory of disordered models of
  glasses.
\newblock {\em Proceedings of the National Academy of Sciences},
  114(13):3328--3333, 2017.

\bibitem{angelini2017real}
Maria~Chiara Angelini and Giulio Biroli.
\newblock Real space migdal--kadanoff renormalisation of glassy systems: recent
  results and a critical assessment.
\newblock {\em Journal of Statistical Physics}, 167(3):476--498, 2017.

\bibitem{Gardner1984}
{Gardner, E.}
\newblock A spin glass model on a hierarchical lattice.
\newblock {\em J. Phys. France}, 45(11):1755--1763, 1984.

\bibitem{angelini2013ensemble}
Maria~Chiara Angelini, Giorgio Parisi, and Federico Ricci-Tersenghi.
\newblock Ensemble renormalization group for disordered systems.
\newblock {\em Phys.\ Rev.\ B}, 87(13):134201, 2013.

\bibitem{dyson1969existence}
Freeman~J Dyson.
\newblock Existence of a phase-transition in a one-dimensional ising
  ferromagnet.
\newblock {\em Communications in Mathematical Physics}, 12(2):91--107, 1969.

\bibitem{meurice2007nonlinear}
Y~Meurice.
\newblock Nonlinear aspects of the renormalization group flows of dyson's
  hierarchical model.
\newblock {\em Journal of Physics A: Mathematical and Theoretical}, 40(23):R39,
  2007.

\bibitem{franz2009overlap}
Silvio Franz, T~J{\"o}rg, and Giorgio Parisi.
\newblock Overlap interfaces in hierarchical spin-glass models.
\newblock {\em Journal of Statistical Mechanics: Theory and Experiment},
  2009(02):P02002, 2009.

\bibitem{angelini2014relations}
Maria~Chiara Angelini, Giorgio Parisi, and Federico Ricci-Tersenghi.
\newblock Relations between short-range and long-range ising models.
\newblock {\em Physical Review E}, 89(6):062120, 2014.

\bibitem{banos2012correspondence}
RA~Banos, LA~Fernandez, V{\'\i}ctor Martin-Mayor, and AP~Young.
\newblock Correspondence between long-range and short-range spin glasses.
\newblock {\em Physical Review B}, 86(13):134416, 2012.

\bibitem{monthus2015fractal}
C{\'e}cile Monthus.
\newblock Fractal dimension of spin-glasses interfaces in dimension d= 2 and d=
  3 via strong disorder renormalization at zero temperature.
\newblock {\em Fractals}, 23(04):1550042, 2015.

\bibitem{wang2018fractal}
Wenlong Wang, MA~Moore, and Helmut~G Katzgraber.
\newblock Fractal dimension of interfaces in edwards-anderson spin glasses for
  up to six space dimensions.
\newblock {\em Physical Review E}, 97(3):032104, 2018.

\bibitem{khoshbakht2018domain}
Hamid Khoshbakht and Martin Weigel.
\newblock Domain-wall excitations in the two-dimensional ising spin glass.
\newblock {\em Physical Review B}, 97(6):064410, 2018.

\bibitem{wang2017number}
Wenlong Wang, Jonathan Machta, Humberto Munoz-Bauza, and Helmut~G Katzgraber.
\newblock Number of thermodynamic states in the three-dimensional
  edwards-anderson spin glass.
\newblock {\em Physical Review B}, 96(18):184417, 2017.

\bibitem{EFETOV1990}
K.B. Efetov.
\newblock Effective medium approximation in the localization theory: Saddle
  point in a lagrangian formulation.
\newblock {\em Physica A: Statistical Mechanics and its Applications},
  167(1):119--131, 1990.

\bibitem{Slanina2006}
Giorgio Parisi and Franti{\v{s}}ek Slanina.
\newblock Loop expansion around the bethe{\textendash}peierls approximation for
  lattice models.
\newblock {\em Journal of Statistical Mechanics: Theory and Experiment},
  2006(02):L02003--L02003, feb 2006.

\bibitem{Sacksteder2007}
Vincent~E. Sacksteder.
\newblock Sums over geometries and improvements on the mean field
  approximation.
\newblock {\em Phys. Rev. D}, 76:105032, Nov 2007.

\bibitem{altieri2017loop}
Ada Altieri, Maria~Chiara Angelini, Carlo Lucibello, Giorgio Parisi, Federico
  Ricci-Tersenghi, and Tommaso Rizzo.
\newblock Loop expansion around the bethe approximation through the m-layer
  construction.
\newblock {\em Journal of Statistical Mechanics: Theory and Experiment},
  2017(11):113303, 2017.

\bibitem{angelini2018one}
Maria~Chiara Angelini, Giorgio Parisi, and Federico Ricci-Tersenghi.
\newblock One-loop topological expansion for spin glasses in the large
  connectivity limit.
\newblock {\em EPL (Europhysics Letters)}, 121(2):27001, 2018.

\bibitem{parisi2014diluted}
Giorgio Parisi, Federico Ricci-Tersenghi, and Tommaso Rizzo.
\newblock Diluted mean-field spin-glass models at criticality.
\newblock {\em Journal of Statistical Mechanics: Theory and Experiment},
  2014(4):P04013, 2014.

\bibitem{angelini2020loop}
Maria~Chiara Angelini, Carlo Lucibello, Giorgio Parisi, Federico
  Ricci-Tersenghi, and Tommaso Rizzo.
\newblock Loop expansion around the bethe solution for the random magnetic
  field ising ferromagnets at zero temperature.
\newblock {\em Proceedings of the National Academy of Sciences},
  117(5):2268--2274, 2020.

\bibitem{angelini2022unexpected}
Maria~Chiara Angelini, Carlo Lucibello, Giorgio Parisi, Gianmarco Perrupato,
  Federico Ricci-Tersenghi, and Tommaso Rizzo.
\newblock Unexpected upper critical dimension for spin glass models in a field
  predicted by the loop expansion around the bethe solution at zero
  temperature.
\newblock {\em Physical Review Letters}, 128(7):075702, 2022.

\end{thebibliography}
\end{document}